\documentclass[rnote,fleqn,oldversion]{aa}
\author{}
\usepackage{natbib,graphicx}
\usepackage{amsmath,amsbsy}
\usepackage[varg]{txfonts}
\usepackage{setspace}

\begin{document}
  \title{Closed-form expressions for particle relative velocities induced by turbulence}
  \author{C. W. Ormel 
          \inst{1} 
          \and 
          J. N. Cuzzi 
          \inst{2}
          }
  \institute{Kapteyn Astronomical Institute, University of Groningen, PO box 800, 9700 AV  Groningen, The Netherlands\\
          \email{ormel@astro.rug.nl}
          \and
          Ames Research Center, NASA, Mail Stop 245-3, Moffett Field, CA 94035, USA\\
          \email{jcuzzi@mail.arc.nasa.gov}
          }
          \abstract{In this note we present complete, closed-form expressions for random relative velocities between colliding particles of arbitrary size in nebula turbulence. These results are exact for very small particles (those with stopping times much shorter than the large eddy overturn time) and are also surprisingly accurate in complete generality (that is, also apply  for particles with stopping times comparable to, or much longer than, the large eddy overturn time). We note that some previous studies may have adopted previous simple expressions, which we find to be in error regarding the size dependence in the large particle regime.}
  \titlerunning{Closed-form expressions for turbulent relative velocities}
  \maketitle
\section{Introduction and outline}
Gas in astrophysical environments is often in a turbulent state of motion,
constantly affected by temporally and spatially varying accelerations from 
eddies having a variety of scales. A particle, due to its inertia, does not
instantaneously follow the gas motions but requires a certain time in order to
align with the gas motion. The particle's interaction with the gas is captured in the definition of the {\it stopping time} of the particle (sometimes also referred to as friction time),
\begin{equation}
  t_\mathrm{s} = \frac{3}{4 c_\mathrm{g} \rho_\mathrm{g}} \frac{m}{\sigma},
  \label{eq:frictime}
\end{equation}
where $c_\mathrm{g}$ and $\rho_\mathrm{g}$ are, respectively, the sound speed and the volume mass density of the gas, and $m$ and $\sigma$ the mass and projected surface area of the particle. Due to this inertial lag, a particle develops a relative
velocity with respect to the gas. In addition, these lags also cause particles to acquire relative velocities among themselves.

While the general problem of calculating these relative velocities has received 
considerable attention in the basic fluid dynamics community (see \citet{2003Icar..164..127C} for references; henceforth CH03), the formalism most frequently used in the astrophysics community 
was developed by \citet{1980A&A....85..316V} and \citet{1991A&A...242..286M} (henceforth MMV). In these works the final results are given in terms of integrals that were not \textit{solved} analytically. Some workers have used simple fits to these numerical results in their models of dust coagulation; however, simple closed-form expressions for
particle-particle relative velocities would help streamline these models \citep[e.g.][]{2001ApJ...551..461S,2005A&A...434..971D,2006ApJ...640.1099N,2007A&A...461..215O}. 
Recently, CH03 obtained closed-form expressions from
the MMV model for particle velocities in inertial space $V_\mathrm{p}$, for particle-gas relative velocities  $V_\mathrm{pg}$, and for relative velocities between two identical particles $V_\mathrm{pp}$, but did not extend their results to the general case of two particles of different stopping times. Moreover, 
CH03 stressed the validity of their analytical results for particles with stopping times much shorter than the large eddy turnover time. In this note we generalize the 
approach and results of CH03 to obtain closed-form
expressions for relative velocities between particles of arbitrary, and unequal, size. 
In Sect.\ \ref{sec:definitions} we define important quantities and review previous work. In Sect.\ \ref{sec:results} we 
present two independent approaches for obtaining the desired closed-form solutions. 
In Sect.\ \ref{sec:conclusions} we give our conclusions and a summary. 

\section{\label{sec:definitions}Definitions and previous work}

Nebula gas turbulence is generally described as being composed of eddies having a 
range of spatial scales $\ell$ and spatial frequencies $k = 1/\ell$, with an energy 
spectrum $E(k) \propto k^{-5/3}$ and total energy $V_\mathrm{g}^2/2$ per unit mass providing 
the normalization condition 
\begin{equation}
  \frac{V_\mathrm{g}^2}{2} = \int_{k_\mathrm{L}}^{k_{\eta}} dk\ E(k),
  \label{eq:normspec}
\end{equation}
from which $E(k) = V_\mathrm{g}^2/3k_\mathrm{L}\ (k/k_\mathrm{L})^{-5/3}$.
The largest, or integral scale, eddies have spatial scale $L = 1/k_\mathrm{L}$, and the smallest, or 
Kolmogorov scale, eddies have spatial scale $\eta = 1/k_{\eta}$. The form of $E(k)$ 
given above is the inertial range expression most often assumed, with $E(k)=0$ for 
$k > k_{\eta}$ or $k < k_\mathrm{L}$. \citet{1980A&A....85..316V} used a spectrum $P(k)=2 E(k)$ 
and stipulated no smallest scale $\eta$ for the turbulence, but \citet{1984Icar...60..553W} 
and MMV noted that a finite value for $\eta > 0$ had profound effects on the 
particle velocities, especially the relative velocities $V_\mathrm{pp}$ for small particles. Each eddy 
wavenumber $k$ has a characteristic velocity $V(k) = \sqrt{2kE(k)}$ and overturn time 
$t_k = \ell/V(k) = (k V(k))^{-1}$. Our standard definition of the particle {\it Stokes number} is $\mathrm{St} = t_\mathrm{s}/t_\mathrm{L}$, where $t_\mathrm{L}$ is the overturn time of the largest eddy, generally 
taken to be the local orbit period. The local turbulent intensity is described by its Reynolds number, $Re$, defined as the ratio between the turbulent and the molecular kinematic viscosities, $Re = \nu_\mathrm{T}/\nu$. The values for $\ell, v$ and $t$ at the integral scale then follow from $Re$, e.g., $\eta = Re^{-3/4} L$ and $t_\eta = Re^{-1/2} t_\mathrm{L}$. These expressions bring $Re$ into the final expressions for particle velocities as a limit on certain integrals ({\it cf.} 
CH03 for more detail). In the notation of astrophysical ``$\alpha$-models", 
$Re = \alpha c_\mathrm{g} H_\mathrm{g} / \nu = \alpha c_\mathrm{g}^2 / \nu \Omega$ where $c_\mathrm{g}$, $H_\mathrm{g}$, and $\nu$ are the sound speed, vertical scale height, and kinematic viscosity of the nebula gas and $\Omega$ is the orbital frequency. 
 
\citet{1980A&A....85..316V} introduced the concept of ``eddy
classes''. Class I eddies vary slowly enough that a particle, upon entering a
class I eddy, will forget its initial motion and align itself to the gas
motions of the eddy before the eddy decays or the particle leaves the eddy.
Class II eddies, on the other hand, have fluctuation times shorter than the 
particle's stopping time $t_\mathrm{s}$, and fluctuate too rapidly to provide more than 
a small perturbation on the
particle. The timescale on which an eddy decays is
given by $t_k$, while the
eddy-crossing timescale is $t_\mathrm{cross} \approx \ell/V_\textrm{rel} = (k
V_\textrm{rel}(k))^{-1}$, with $V_\textrm{rel}$ the relative velocity between 
a grain and an eddy. For an eddy
to be of class I both $t_k$ and $t_\mathrm{cross}$ must be larger than the
particle's stopping time. The boundary between these classes occurs at $k=k^*$ (or at $t_k = t^*)$ which can be defined as
(\citet{1980A&A....85..316V}, MMV):
\begin{equation}
  \frac{1}{t_\mathrm{s}} = \frac{1}{t^*} + \frac{1}{t_\mathrm{cross}}
                = \frac{1}{t^*} + k^* V_\textrm{rel}(k^*).
  \label{eq:tstardef}
\end{equation}
It is important to realize that $k^*$ (or $t^*$) is a function of stopping time $t_\mathrm{s}$, that is, the boundary separating the two classes is different for each particle.
The different treatment for the two eddy classes $k<k^*$ and $k>k^*$ forms the 
core of the derivation of the turbulence-induced particle velocities. 

All turbulent velocities in this note are statistical, root-mean-square, averaged quantities. The average inertial space particle velocity $V_\mathrm{p}$ is given by
Eq.\ (6) of MMV.
\begin{align}
  \nonumber
  V_\mathrm{p}^2 =&\ \int_{k_\mathrm{L}}^{\texttt{max}(k^*,k_\mathrm{L})} dk\ 2E(k) \left( 1 - K^2 \right) \\
     &\ + \int_{\texttt{max}(k^*,k_\mathrm{L})}^{k_{\eta}} dk\ 2E(k) \left( 1-K \right)
\left[ g(\chi) + K h(\chi) \right],
  \label{eq:Vp0}
\end{align}
in which $K = t_\mathrm{s}/(t_\mathrm{s}+t_k)$. The $K^2$ term in the first integral results from the more recently preferred ``$n=1$" gas velocity autocorrelation function (MMV and CH03). The functions $g(\chi)=\chi^{-1}{\rm tan}^{-1}(\chi)$ and $h(\chi) = 1/(1+\chi^2)$ with  $\chi = K t_k k V_\textrm{rel}$ were first obtained by \citet{1980A&A....85..316V}. 

CH03 noted that, for very small particles with $t_\mathrm{s} \ll t_\mathrm{L}$
or $\mathrm{St} \ll 1$, the second integral becomes negligible, leaving only the first integral which is analytically solvable and for which the upper limit can be extended to $k_{\eta}$ with negligible error.
Here,  to generalize the approach of CH03 to particles of {\it arbitrary} size, we approximate $h(\chi) = g(\chi) = 1$ for {\it all} particle sizes (see CH03 Sect.\ 2.2.3 for supporting logic). Numerical calculations of $h(\chi)$ and $g(\chi)$ validate this approximation to order unity (see Appendix\ \ref{app:app1}), and we gain further confidence in it from {\it a posteriori} comparison with exact numerical model results. The general expression for $V_\mathrm{p}^2$ is then the same as in the $t_\mathrm{s} \ll t_\mathrm{L}$ regime, and the same analytical result is obtained, {\it i.e.}
CH03,
\begin{align}
  \label{eq:Vp1}
  V_\mathrm{p}^2 =&\ \int_{k_\mathrm{L}}^{k_{\eta}} dk\ 2E(k) \left( 1-K^2 \right) \\
  =&\ V_\mathrm{g}^2 \left( 1 -
\frac{\mathrm{St}^2(1-Re^{-1/2})}{(\mathrm{St}+1)(\mathrm{St}+Re^{-1/2})}
\right).
  \label{eq:Vp2}
\end{align}
CH03 did not give this explicit result for $V_\mathrm{p}$, but merely noted that it was 
straightforward to derive it from their Eq.\ (19) for $V_\mathrm{pg}$ and the general 
relationship $V_\mathrm{pg}^2 = V_\mathrm{p}^2 - V_\mathrm{g}^2$; however we will use it explicitly here. 

Comparison of the predictions of this simple expression with detailed numerical 
results (MMV, CH03) show that it is indeed a good approximation for arbitrary 
$\mathrm{St}$. A more accurate approximation to Eq.\ (\ref{eq:Vp0}), in which the $g$ and $h$ functions are approximated as power-laws in $k^*/k$, is outlined in Appendix\ \ref{app:app1}. Unless $\mathrm{St} \ll 1$, we can neglect 
the Reynolds number term in Eq.\ (\ref{eq:Vp2}) and obtain 
$V_\mathrm{p} = V_\mathrm{g}/\sqrt{1+\mathrm{St}}$, a well known result \citep{1980A&A....85..316V,1993Icar..106..102C,2004ApJ...614..960S} which describes the diffusivity of large particles in turbulence. 

\section{\label{sec:results}Results}
\subsection{\label{sec:kspace}k-space approach}
MMV (their Eq.\ (7)) expressed the relative velocities $V_\mathrm{p1p2}$ between particles of different stopping times $t_1$ and $t_2$ as 
  \begin{equation}
V_\mathrm{p1p2}^2 = V_\mathrm{p1}^2 +V_\mathrm{p2}^2 - 2 \overline{V_\mathrm{p1}V_\mathrm{p2}} \equiv \Delta V_{12}^2. 
  \end{equation}
  Having already derived $V_{\mathrm{p}i}^2$ ($i=1,2$) above, we can determine $\Delta V_{12}$ by evaluating the cross term $\overline{V_\mathrm{p1}V_\mathrm{p2}}$; this paper presents analytical solutions of this problem obtained in two separate ways. In this subsection we retain the wavenumber dependence; in the next subsection we transform to time variables. In Eq. (8) of MMV the cross term is given as a sum over the two particle sizes involved, which we separate here, writing $\Delta V_{12}^2  = V_\mathrm{p1}^2 + V_\mathrm{p2}^2 - (V_\mathrm{c1}^2 + V_\mathrm{c2}^2)$, where 

\begin{multline}
  V_{\mathrm{c}i}^2 = \frac{2t_i}{t_1+t_2} \left( \int_{k_\mathrm{L}}^{\texttt{min}(k_1^*,k_2^*)} E(k) dk \right. \\   
\left. - \int_{k_\mathrm{L}}^{\texttt{min}(k_1^*,k_2^*)} E(k) \left(  \frac{1}{1 + t_k/t_i }\right)^2 dk \right).
  \label{eq:Vcross}
\end{multline}
Changing variable to $x=k/k_\mathrm{L}$, substituting for $E(k)$, and converting stopping time $t_i$ to Stokes number $\mathrm{St}_i = t_i/t_\mathrm{L}$:
\begin{equation}
  V_{\mathrm{c}i}^2 = \frac{2V_\mathrm{g}^2 t_i}{3(t_1+t_2)} \left[ \int_1^{x_1^*} x^{-5/3}dx 
 - \int_1^{x_1^*} \frac{\mathrm{St}_i^2 dx}{x^{5/3}(\mathrm{St}_i + x^{-2/3})^2} \right],
\end{equation}
where we have taken, without loss of generality, $k_1^* \le k_2^*$. The first integral is trivial and the second integral can be solved exactly as in Eqs.\ (17--19) of CH03. In evaluating the specific value of the integrals above, we need a closed form for the upper limit $x_1^* = k_1^*/k_\mathrm{L}$. A simple prescription is readily found by inspection of Fig.\ 3 of CH03: $x_1^* = k_1^*/k_\mathrm{L} = 0.5\mathrm{St}_1^{-3/2} +1$. That is, the boundary eddy for particles with stopping time $t_1$ is that for which $t_k \sim t_1$ until $t_1 > t_\mathrm{L}$, beyond which it remains constant. This is merely a convenient mathematical shorthand to keep everything in closed form. Then, repeating the analytical solution of CH03 (Eqs.\ (17--19)) we obtain 
\begin{multline}
  V_{\mathrm{c}i}^2 = V_\mathrm{g}^2 \frac{t_i}{(t_1+t_2)} \left[ (1 - {x_1^*}^{-2/3} ) - \right. \\
\left. \left( \frac{\mathrm{St}_i}{1 + \mathrm{St}_i}  - \frac{\mathrm{St}_i}{1 + \mathrm{St}_i {x_1^*}^{2/3} } \right) \right].
\end{multline}

This solution for the cross term is easily combined with \mbox{Eq.\ (\ref{eq:Vp2})} to obtain expressions for particle-particle relative velocities $\Delta V_{12}^2$. Further manipulation of these expressions may be possible, but the important point here is that $\Delta V_{12}$ can be expressed in closed form as function of $\mathrm{St}_1$,  $\mathrm{St}_2$, $V_\mathrm{g}$, and $Re$. With a few minutes of algebra, simpler expressions can be found in the limiting regimes of interest ($\mathrm{St}_1 \ll 1, \gg 1,$ {\it etc.}) which agree well with those which we present in the next section, for analytical solutions obtained in the time domain instead of the wavenumber domain, and where an analytical solution for the boundary $k^*(t^*)$ is used rather than the form for $x_1^*$ adopted above. It should be recalled that, for {\it very} small particles $t_\mathrm{s} < t_\eta$, $x_1^*$ has an upper limit of $k_\eta/k_\mathrm{L} = Re^{3/4}$ (see, {\it e.g.} CH03 Fig.\ 3).  

\subsection{t-space approach}
The integrals expressing $V_{\mathrm{p}i}^2$ and $V^2_{\mathrm{c}i}$ are transformed into a simpler
 form by changing variables from $k$ to $t_k$. Since $t_k = 1/k V(k) = 
\left(k \sqrt{2kE(k)}\right)^{-1}$ and $E(k) = A k^{-5/3}$ for a Kolmogorov
power spectrum (where $A$ is the normalization factor), we obtain that $E(k) dk =
\frac{3}{2} \sqrt{2} A^{3/2} dt_k$. Now, $A=\tfrac{1}{3}V_\mathrm{g}^2 k_\mathrm{L}^{2/3}$ from
the normalization of the turbulent spectrum (Eq.\ (\ref{eq:normspec})), $k_\mathrm{L} =
(V_\mathrm{L} t_\mathrm{L})^{-1}$ with $V_\mathrm{L}$ the velocity of the largest eddy, and $V_\mathrm{L}^2 =
\frac{2}{3} V_\mathrm{g}^2$ also by normalizing the power spectrum (see CH03). 
We then end up with
\begin{equation}
  E(k)\ dk = \frac{1}{2} \frac{V_\mathrm{g}^2}{t_\mathrm{L}} dt_k,
  \label{eq:powerspectrum}
\end{equation}
which can be substituted into all the integrals, putting them into a simpler form. 
For instance, Eq.\ (\ref{eq:Vp1}) becomes for particle $i$ 
\begin{equation}
  V_{\mathrm{p}i}^2 = \frac{V_\mathrm{g}^2}{t_\mathrm{L}} \int_{t_\eta}^{t_\mathrm{L}} dt_k \left( 1 - \left(
\frac{t_i}{t_i+t_k} \right)^2 \right) = \frac{V_\mathrm{g}^2}{t_\mathrm{L}} \left[ t_k +
\frac{t_i^2}{t_i + t_k} \right]^{t_\mathrm{L}}_{t_\eta}.\\
  \label{eq:Vp3}
\end{equation}
Similarly, the cross term becomes
\begin{align}
  V_{\mathrm{c}i}^2 =& \frac{V_\mathrm{g}^2}{t_\mathrm{L}} \frac{2t_i}{t_1+t_2} \int_{t^*_{12}}^{t_\mathrm{L}}
dt_k\ \left( 1 - \left( \frac{t_i}{t_i + t_k} \right)^2 \right) \\   =&\
\frac{V_\mathrm{g}^2}{t_\mathrm{L}} \frac{2t_i}{t_1+t_2}\left[ t_k + \frac{t_i^2}{t_i + t_k}
\right]^{t_\mathrm{L}}_{t_{12}^*}
  \label{eq:vci}
\end{align}
With $t^*_{12} = \texttt{max}(t_1^*,t_2^*$) and $t_\eta \le t_{12}^* \le t_\mathrm{L}$ since $t^*$ refers to an eddy's turn-over time. We now solve for $\Delta V_{12}^2$ by splitting
the integral in Eq.\ (\ref{eq:Vp3}) at $t^*_{12}$ and subtracting the corresponding 
$V_{\mathrm{c}i}$ terms from Eq.\ (\ref{eq:vci}) to get
\begin{multline}
  \Delta V_{12}^2 = \frac{V_\mathrm{g}^2}{t_\mathrm{L}} \left( \left[ t_k+\frac{t_1^2}{t_1+t_k}
\right]_{t_\eta}^{t_{12}^*} + \left[ t_k+\frac{t_1^2}{t_1+t_k}
\right]_{t^*_{12}}^{t_\mathrm{L}} - \right. \\   \left. \frac{2t_1}{t_1+t_2}\left[ t_k +
\frac{t_1^2}{t_1 + t_k} \right]^{t_\mathrm{L}}_{t_{12}^*}  + (1\leftrightarrow2)
\right),
\end{multline}
where the $(1\leftrightarrow2)$ symbol indicates interchange between particles 1 and 2. With further manipulation and cancellation
of terms, the previous expression simplifies slightly to
\begin{align}
  \Delta V_{12}^2 =&\ \frac{V_\mathrm{g}^2}{t_\mathrm{L}} \left(  \left[ t_k +\frac{t_1^2}{t_1+ t_k}
\right]_{t_\eta}^{t_{12}^*}
+ \frac{t_2-t_1}{t_1+t_2}  \left[ \frac{t_1^2}{t_1 +t_k}
  \label{eq:delv}
\right]^{t_\mathrm{L}}_{t_{12}^*} + (1\leftrightarrow2) \right) \\
  \nonumber
  =&\ \Delta V_\mathrm{II}^2 + \Delta V_\mathrm{I}^2
\end{align}
This is perhaps the most concise way to write the expressions for $\Delta
V_{12}^2$. The first term we call $\Delta V_\mathrm{II}^2$ since this term involves class
II (fast) eddies. If $t_{12}^* = t_\mathrm{L}$ (heavy particles) all eddies are fast and only this term
remains. Conversely, if $t_{12}^* = t_\eta$ (small particles) the contribution from $\Delta V_\mathrm{II}$
vanishes and the second term, $\Delta V_\mathrm{I}$, determines relative velocities. In
the intermediate regime, $t_\eta < t_{12}^* < t_\mathrm{L}$, both terms contribute. Written
in terms of the Stokes numbers these terms becomes
\begin{align}
  \nonumber
  \Delta V_\mathrm{I}^2 \equiv &\ \frac{V_\mathrm{g}^2}{t_\mathrm{L}} \frac{t_2-t_1}{t_1+t_2}  \left[ \frac{t_1^2}{t_1 +t_k} \right]^{t_\mathrm{L}}_{t_{12}^*} + (1\leftrightarrow2) \\
  =&\ V_\mathrm{g}^2 \frac{\mathrm{St}_1 - \mathrm{St}_2}{\mathrm{St}_1+\mathrm{St}_2} \left( \frac{\mathrm{St}_1^2}{\mathrm{St}_{12}^*+\mathrm{St}_1} - \frac{\mathrm{St}_1^2}{1+\mathrm{St}_1} - (1\leftrightarrow2) \right)
  \label{eq:delvI}
\end{align}
\begin{multline}
  \Delta V_\mathrm{II}^2 \equiv \frac{V_\mathrm{g}^2}{t_\mathrm{L}} \left[ t_k +\frac{t_1^2}{t_1+ t_k}
\right]_{t_\eta}^{t_{12}^*} + (1\leftrightarrow2)  \\
  \hspace{-5mm}
  = V_\mathrm{g}^2 \left( (\mathrm{St}_{12}^* - Re^{-1/2}) +
   \frac{\mathrm{St}_1^2}{\mathrm{St}_1+\mathrm{St}_{12}^*} - \frac{\mathrm{St}_1^2}{\mathrm{St}_1+Re^{-1/2}} + (1\leftrightarrow2) \right)
  \label{eq:delvII}
\end{multline}
Note again that since $t_\eta \le t^*_{12} \le t_\mathrm{L}$ we also have that $Re^{-1/2} \le
\mathrm{St}_{12}^* \le 1$. Below, we will first solve
for $\mathrm{St}^*_{12}$, and then consider solutions for $\Delta V_{12}$ in various  
limiting cases of the particle stopping times.

\subsubsection{\label{sec:tstar}Solving for $t^*$}
The relative velocity between a particle with stopping time $t_\mathrm{s}$ and an eddy
$k$, is given by \citet{1980A&A....85..316V}, Eq. (15):
\begin{equation}
  V_\textrm{rel}(k)^2 = V_\mathrm{o}^2 + 2\int_{k_\mathrm{L}}^{k} dk' E(k') \left( \frac{t_\mathrm{s}}{t_\mathrm{s}+t_k} \right)^2.
  \label{eq:Vreldef}
\end{equation}
$V_\mathrm{o}$ is any systematic velocity component not driven by turbulence -- such as due to pressure-gradient driven azimuthal headwind, the ensuing radial drift, or vertical settling under solar gravity. We can integrate this equation in the same fashion as Eq.\ (\ref{eq:vci}) and arrive at
\begin{equation}
  V_\textrm{rel}(k^*)^2 = V_\mathrm{o}^2 + \frac{V_\mathrm{g}^2}{t_\mathrm{L}} \left[
\frac{t_\mathrm{s}^2}{t_\mathrm{s}+t_k} \right]^{t^*}_{t_\mathrm{L}} \\
= V_\mathrm{o}^2 + \frac{V_\mathrm{g}^2
t_\mathrm{s}}{t_\mathrm{L}} \left( \frac{1}{1+y^*} - \frac{1}{1+y_\mathrm{L}} \right),
\label{eq:vreldef2}
\end{equation}
in which $y=t_k/t_\mathrm{s}$. Also, using the definition for $t_k$ (see text above Eq.\ (\ref{eq:powerspectrum})), $k^*$ can be expressed as  
$({k^*})^2=(2A)^{-3/2} {t^*}^{-3} = \tfrac{3}{2} V_\mathrm{g}^{-2} t_\mathrm{L} {t^*}^{-3}$.
Inserting the expressions for $k^*$ and $V_\mathrm{rel}^2$ into Eq.\ (\ref{eq:tstardef}), assuming that $V_\textrm{o} =0$ for simplicity (see however Sect.\ \ref{sec:roleofV0}), we obtain:
\begin{figure}[t]
  \includegraphics[width=88mm]{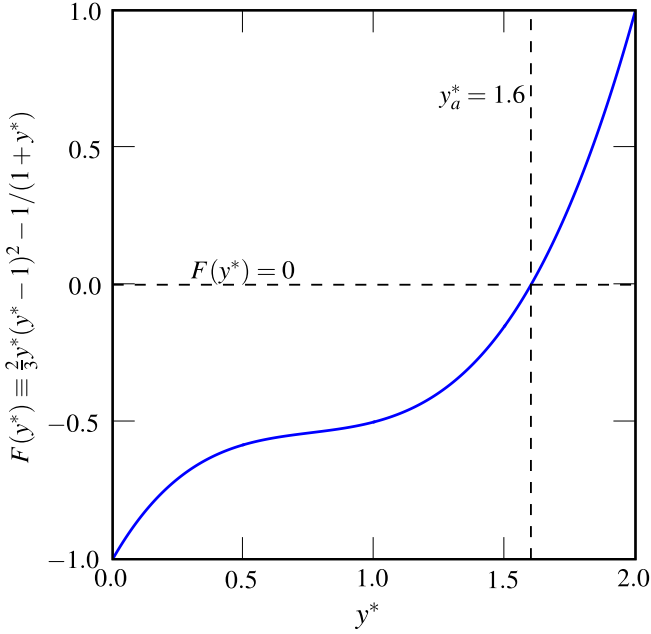}
  \caption{The function $\tfrac{2}{3}y^*(y^*-1)^2 - 1/(1+y^*)$. If $t_\mathrm{s} \ll t_\mathrm{L}$ and 
$V_\mathrm{o}=0$ (no systematic velocity drifts; see Sect.\ \ref{sec:roleofV0}) this equation is equal to zero 
and we find a solution $y^* = t^*/t_\mathrm{s} \approx y_\mathrm{a}^* = 1.6$. On the other hand, for $t_\mathrm{s} \sim t_\mathrm{L}$, 
the RHS of Eq.\ (\ref{eq:ystareq}) is $\approx -0.5$ and $y^* \approx 1$.}
  \label{fig:function}
\end{figure}
\begin{figure}[t!]
  \includegraphics[width=88mm]{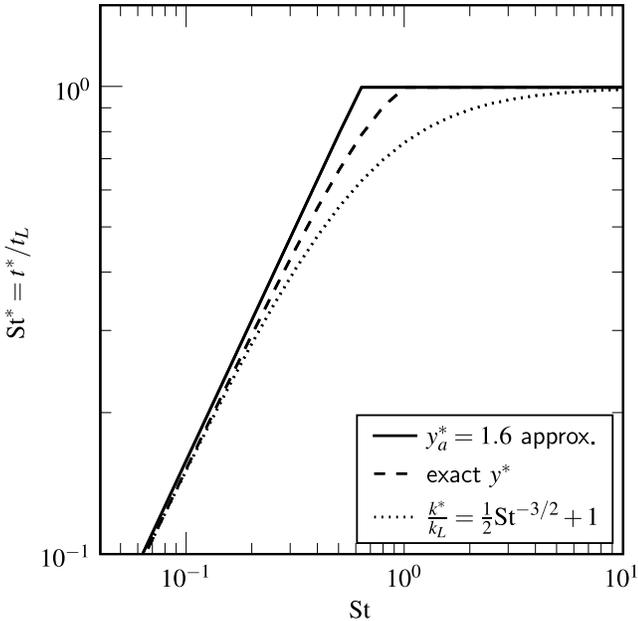}
  \caption{Three different assumptions for $t^*$ (or the related $k^*$) are shown here.}
  \label{fig:tstarplot}
\end{figure}
\begin{subequations}
\begin{equation}
  \frac{1}{k^*}\left(\frac{1}{t_\mathrm{s}} - \frac{1}{t^*}\right) = V_\textrm{rel} \Rightarrow\\
\end{equation}
\begin{equation}
  t{^*}\left( \frac{t^*}{t_\mathrm{s}} - \frac{t^*}{t^*} \right)^2 = (2A)^{-3/2} V^2_\textrm{rel} \Rightarrow\\
\end{equation}
\begin{equation}
  t^*\left( y^* -1 \right)^2 = \tfrac{3}{2} t_\mathrm{s}\left( \frac{1}{1+y^*} -
\frac{1}{1+y_\mathrm{L}} \right) \Rightarrow\\
\end{equation}
\begin{equation}
  \tfrac{2}{3} y^* \left( y^* -1 \right)^2 - \frac{1}{1 + y^*} = - \frac{1}{1+y_\mathrm{L}},
  \label{eq:ystareq}
\end{equation}
\end{subequations}
where we have defined $y^* = t^*/t_\mathrm{s}$ and $y_\mathrm{L} = t_\mathrm{L}/t_\mathrm{s} = \mathrm{St}^{-1}$.
The LHS of Eq.\ (\ref{eq:ystareq}) is plotted in Fig.\ \ref{fig:function}. If $y_\mathrm{L} \gg 1$ for 
small particles, the RHS of Eq.\ (\ref{eq:ystareq}) is negligible and the numerical solution for $y^*$ becomes
$y^* \approx y_a^* = 1.6$, or  $t^* \approx 1.6 t_\mathrm{s}$. On the other hand, when $t_\mathrm{s}$ nears $t_\mathrm{L}$, the $-1/(1+y_\mathrm{L})$
term causes the RHS of Eq.\ (\ref{eq:ystareq}) to drop to $-0.5$,
and $y^* \rightarrow 1$. For $t_\mathrm{s} > t_\mathrm{L}$ we always have that $t^*=t_\mathrm{L}$; 
{\it i.e.}, for such a particle all eddies
are of class 2. In Fig.\ \ref{fig:tstarplot} we compare the exact solution (dashed line)
for $t^*$ with the $t^*\approx y^*_a t_\mathrm{s} = 1.6t_\mathrm{s}$ approximation (solid line; in both cases $t^* \le t_\mathrm{L}$
is simply enforced), and the empirical function
$k^*/k_\mathrm{L} = 1 + \tfrac{1}{2}\mathrm{St}^{-3/2}$ (dotted line; see Sect.\ \ref{sec:kspace}).

The exact solution for $\Delta V_{12}$ (Eq.\ \ref{eq:delv}) is given in Fig.\
\ref{fig:delv} both for $t_1 \gg t_2$ (solid curve) and for particles of equal
stopping times (dashed curve). A Reynolds number of $Re=10^8$ has been adopted.

\subsection{\label{sec:roleofV0}The role of $V_\mathrm{o}$: eddy-crossing effects}
Systematic velocities $V_\mathrm{o}$ due to vertical settling, and pressure-gradient headwinds and drifts, will occur ({\it eg.} \citet{1986Icar...67..375N}). Because particles drift through eddies, their transit time is affected (because $V_\textrm{rel}$ is larger) and the boundary between class I and II eddies shifts. \citet{1993Icar..106..102C} include this effect, due to vertical settling, in their model of particle diffusion (their Eq.\ (43)). The model presented here offers a generalized way of treating this effect, which we will only sketch here. 

Repeating the procedure outlined in Sect.\ \ref{sec:tstar} but retaining the $V_0$ term in $V_\textrm{rel}$ (Eq.\ (\ref{eq:vreldef2})), we end up with Eq.\ (\ref{eq:ystareq}) including a correction term

\begin{equation}
\tfrac{2}{3} y^* \left( y^* -1 \right)^2 - \frac{1}{1 + y^*} \equiv F(y^*) =  - \frac{\mathrm{St}}{1+\mathrm{St}} + \frac{1}{\mathrm{St}} \frac{V^2_\mathrm{o}}{V^2_g},
\label{eq:voeffect}
\end{equation}
where we have substituted $\mathrm{St} = 1/y_\mathrm{L}$. The correction term can be roughly constrained using an estimate of the systematic drift velocity 
$V_\mathrm{o} \sim (\mathrm{St}/(\mathrm{St} + 1)) \beta V_K$, where $V_\mathrm{K}$ is the Keplerian velocity 
at distance $a$ from the Sun, $\Omega$ is the orbit frequency, and $\beta = (H_\mathrm{g}/a)^2$ is a radial pressure gradient parameter; also we take $V_\mathrm{g} = \alpha^{1/2}c_\mathrm{g}$ (see, {\it eg.}, \citet{1986Icar...67..375N} or \citet{2006mess.book..353C}). Then
\begin{align}
  \nonumber
  \frac{V_\mathrm{o}}{V_\mathrm{g}}=&\ \frac{\mathrm{St}}{\mathrm{St} + 1} \frac{\beta V_K}{\alpha^{1/2}c_\mathrm{g}}
  =  \frac{\mathrm{St}}{\mathrm{St} + 1} \frac{\beta a \Omega}{\alpha^{1/2} H_\mathrm{g} \Omega} \\
  =&\ \frac{\mathrm{St}}{\mathrm{St} + 1} \frac{\beta }{\alpha^{1/2} \beta^{1/2}}
  = \frac{\mathrm{St}}{\mathrm{St} + 1} \left( \frac{\beta}{\alpha} \right)^{1/2},
\end{align}
and Eq.\ (\ref{eq:voeffect}) becomes,
\begin{equation}
  F(y^*) = \frac{\mathrm{St}}{1+\mathrm{St}} \left( \frac{\beta / \alpha}{1+\mathrm{St}} - 1 \right).
  \label{eq:Fystar}
\end{equation}
Normally $\beta \sim 2 \times 10^{-3}$ is assumed \citep{1986Icar...67..375N,1993Icar..106..102C}, but its real value, and that of $\alpha$, are not well known. Equation\ (\ref{eq:Fystar}) shows that for a given value of $\mathrm{St}$, $F(y^*)$ increases with increasing $\beta/\alpha$. Consequently, $y^* = t^*/t_\mathrm{s}$ is also higher (see Fig.\ \ref{fig:function}). The boundary between the class I and II eddies therefore shifts to higher values of $t^*$, that is, there are less class I eddies for high $\beta/\alpha$ and the $\mathrm{St}^*=1$ upper limit (when $t^*=t_\mathrm{L})$ is reached at lower Stokes numbers. Inserting the definition of $F(y^*)$ (LHS of Eq.\ (\ref{eq:voeffect})) into Eq.\ (\ref{eq:Fystar}) with $y^* = t_\mathrm{L}/t_\mathrm{s} = \mathrm{St}^{-1}$ and solving for $\mathrm{St}$, we find that the Stokes number at which $\mathrm{St}^*=1$ occurs at
\begin{equation}
  \mathrm{St}_{\mathrm{St}^*=1} = \left( 1+\sqrt{\frac{3\beta}{2\alpha}} \right)^{-1/2}.
\end{equation}
For example, for $\beta/\alpha =1$, $\mathrm{St}^*$ reaches its upper limit at $\mathrm{St} \approx 0.67$.

In the small particle regime ($\mathrm{St} \ll 1$), however, the exact value of $\beta/\alpha$ is unimportant since $F(y^*)$ is always close to zero, and the $y_\mathrm{a}^*$ approximation is justified. It is only for $\beta/\alpha \gtrsim \mathrm{St}^{-1}$ that the RHS of Eq.\ (\ref{eq:Fystar}) starts to becomes significant and $y^*>y^*_\mathrm{a}$. This is the weakly-turbulent or non-turbulent regime where class II eddies dominate even for small particles. In practise, however, it means that eddy crossing effects are important only if turbulence is very weak and we will not treat them 
further in this paper.

\subsection{Limiting solutions}
\begin{figure}[t]
  \includegraphics[width=88mm]{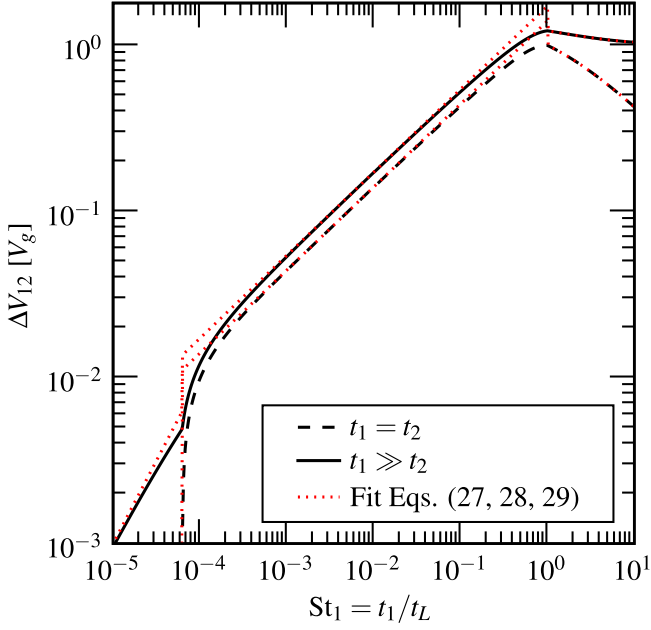}
  \caption{Exact solution to Eq.\ (\ref{eq:delv}) for $\Delta V_{12}$ in
the case of identical particles (dashed line) and $t_1 \gg t_2$ (solid line) for a Reynold number of $10^{8}$.
The dotted curves are approximations to Eq.\ (\ref{eq:delv}) given by \mbox{Eqs.\ (\ref{eq:verysmalllim}, \ref{eq:fullintermlim}, \ref{eq:largelim})}.}
  \label{fig:delv}
\end{figure}
As intuition-building examples we obtain simple, closed-form expressions for $\Delta V_{12}^2$ in various limiting 
regimes from the $t$-space solutions; similar results are easily obtained from the $k$-space 
solutions (Sect.\ \ref{sec:kspace}). Without loss of generality we take particle 1 to have the largest
stopping time, {\it i.e.}, $t_1 \ge t_2$ and $t_{12}^*=t_1^*$. Moreover, we assume that
$t_\eta \ll t_\mathrm{L}$; {\it i.e.}, that $Re^{1/2} \gg 1$ and there is an extended inertial 
range of eddies. Recall again that $\mathrm{St}_{12}^*=Re^{-1/2}$ for $t_1 < t_\eta/y_\mathrm{a}^*$, and that $\mathrm{St}_{12}^*$ will not exceed 1.

\subsubsection{Tightly coupled particles, $t_1,t_2 < t_\eta$}
In this limit all eddies are of class I and $\Delta V_{12}^2 \rightarrow \Delta V_\mathrm{I}^2$. 
For each particle, the second term on the RHS of Eq.\ (\ref{eq:delvI}) is negligible; thus 
\begin{equation}
  \Delta V_{12}^2 = V_\mathrm{g}^2
\frac{\mathrm{St}_1-\mathrm{St}_2}{\mathrm{St}_1+\mathrm{St}_2} \left(
\frac{\mathrm{St}_1^2}{\mathrm{St}_1+Re^{-1/2}} - 
\frac{\mathrm{St}_2^2}{\mathrm{St}_2+Re^{-1/2}} \right).
  \label{eq:smalllim}
\end{equation}
In the very small particle regime ($t_1 \ll t_\eta$), $\mathrm{St}_i \ll t_\eta/t_\mathrm{L} = Re^{-1/2} $ and
\begin{equation}
  \Delta V_{12}^2 = V_\mathrm{g}^2 \frac{t_\mathrm{L}}{t_\eta} \left( \mathrm{St}_1 - \mathrm{St}_2 \right)^2.
  \label{eq:verysmalllim}
\end{equation}
Since $V_\mathrm{g}^2 = \frac{3}{2}V_\eta^2 Re^{1/2} = \tfrac{3}{2}V_\eta^2 t_\mathrm{L}/t_\eta$, this
expression transforms directly to $\Delta V_{12} = \sqrt{3/2} (t_1 - t_2)V_\eta/t_\eta$, in
good agreement with the heuristic, although physically motivated, expression
$\Delta V_{12} = V_\eta (t_1 - t_2)/t_\eta$ of \citet{1984Icar...60..553W}. 

\subsubsection{Intermediate regime, $t_\eta \le t_1 \le t_\mathrm{L}$.}
If $t_1$ (the stopping time of the larger particle) approaches the Kolmogorov 
scale, two changes
occur. First, the $\mathrm{St}_1^2/(\mathrm{St}^*_{12} + \mathrm{St}_1)$ term
in Eq.\ (\ref{eq:delvI}) now becomes linear with $\mathrm{St}_1$, since
$\mathrm{St}_{12}^*$ grows proportional to $\mathrm{St}_1$ (the second term is
still negligible throughout most of this regime). Relative velocities therefore
increase as the square-root of stopping time. Second, class II eddies also
contribute to $\Delta V_{12}^2$ (Eq.\ \ref{eq:delvII}). This contribution scales
also with $\mathrm{St}_1$, but is significantly larger and does not
disappear when $t_1 = t_2$. From a physical point of view, class II eddies act
as small, random kicks to the particle trajectory, while two
particles captured by a class I eddy are subject to the same,
systematic, change in motion. Class II eddies are therefore much more effective
in generating velocity differences for similar-sized particles.

In the ``fully intermediate regime'', i.e., $t_\eta \ll t_1 \ll t_\mathrm{L}$, we can also
ignore the $Re^{-1/2}$ terms in Eq.\ (\ref{eq:delvII}). In addition, the $t^*/t_\mathrm{s} = y_\mathrm{a}^*$ approximation holds. Upon writing $\mathrm{St}_2 = \epsilon \mathrm{St}_1$, Eqs.\ (\ref{eq:delvI}, \ref{eq:delvII}) become linear with $\mathrm{St}_1$ and we can write $\Delta V_{12}^2$ as (see Appendix \ref{ap:app2})
\begin{equation}
  \Delta V_{12}^2 = V_\mathrm{g}^2 \left[ 2y_\mathrm{a} - (1+\epsilon) + \frac{2}{1+\epsilon}\left(
\frac{1}{1+y_\mathrm{a}} + \frac{\epsilon^3}{y_\mathrm{a}+\epsilon} \right) \right]
\mathrm{St}_1,
  \label{eq:fullintermlim}
\end{equation}
where $\epsilon \le 1$ is the ratio between the stopping times and $y_\mathrm{a} = 1.6$. For $t_1 \gg t_2$ we then find that
$\Delta V_{12}^2 \approx 3.0 V_\mathrm{g}^2 \mathrm{St}_1$, while for equal
particles the numerical factor goes down to $2.0$. Written in terms of stopping
times the relative velocities become, $\Delta V_{12} = [1.7\div2.1] V_\mathrm{L}
\sqrt{t_1/t_\mathrm{L}}$. This also compares well with \citet{1984Icar...60..553W} fits
for this regime (who gives pre-factors of $2.1$ and $3.0$, respectively). Note,
however, that our full expressions for $\Delta V$ (Eqs.\ \ref{eq:delv},
\ref{eq:delvI}, \ref{eq:delvII}) also capture the behavior near the $t_\eta$ and
$t_\mathrm{L}$ ``turning points'' (see Fig.\ \ref{fig:delv}).

\subsubsection{Heavy particles, $t_1 > t_\mathrm{L}$}
\begin{figure*}[t]
  \centering
  \includegraphics[width=140mm]{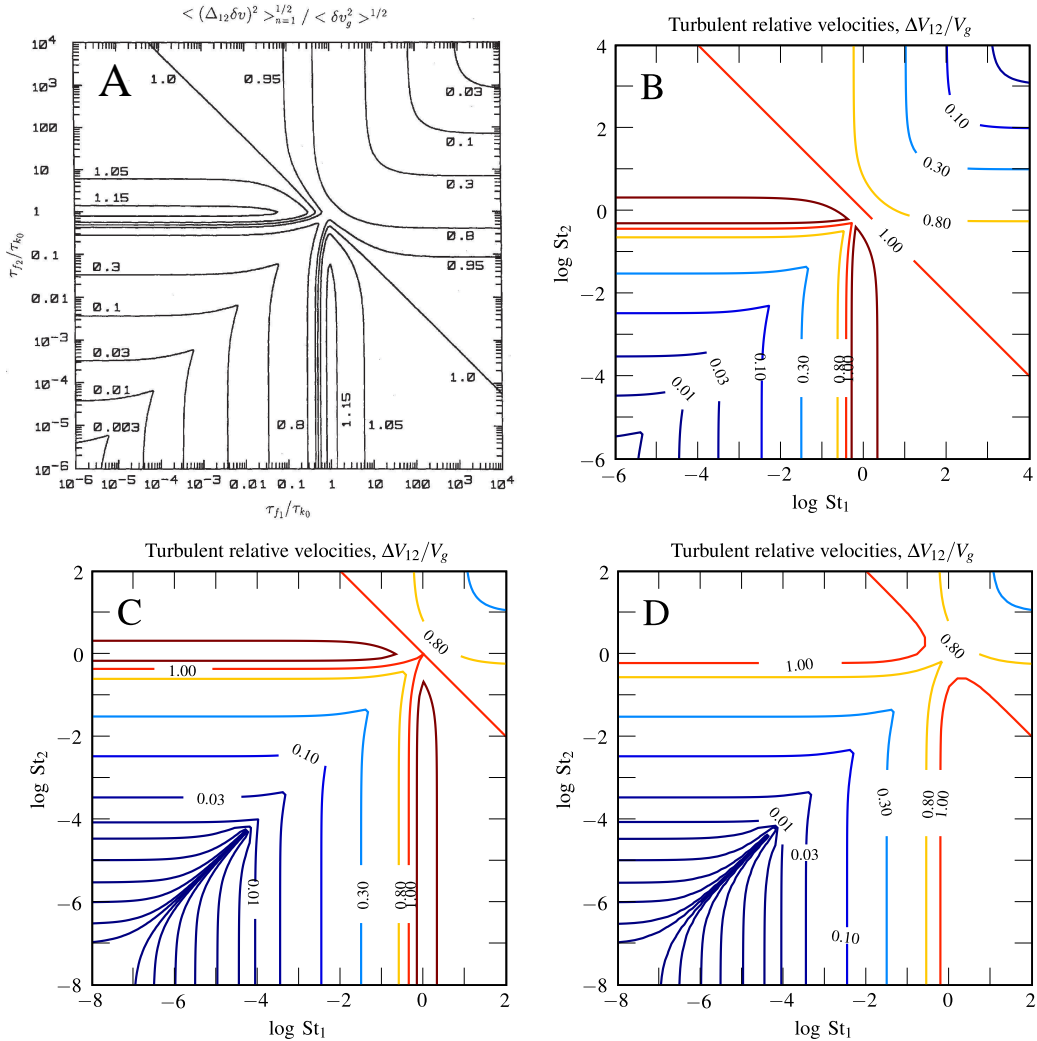}
  \caption{Contour plots of particle-particle, turbulence induced, relative velocities $\Delta V_{12}$
  normalized to $V_\mathrm{g}$. --A-- Numerical results of \citet{1991A&A...242..286M}, without inner scale ($Re\rightarrow \infty$). --B-- Analogous result from our closed-form expressions with the fixed $y^* \approx y^*_\mathrm{a} = 1.6$ approximation {(Sect.\ \ref{sec:tstar})}. --C-- Like B, but with an exact solution for $y^*$ and with $Re=10^8$. --D-- Using the CH03 formula for $k^*$, $k^*/k_L = 0.5\mathrm{St}^* + 1$, and also with $Re=10^8$. Contours are drawn twice per logarithmic decade (at $\Delta V_{12}/V_\mathrm{g} = 3\times 10^i$ and at $10^i$) with an additional contour at $0.8$ and $1.15$.}
  \label{fig:contour}
\end{figure*}
If $t_1 > t_\mathrm{L}$, $\mathrm{St}_{12}^*=1$ and there is no contribution from class
I eddies (Eq.\ \ref{eq:delvI}). Also, we can neglect the $Re^{-1/2}$ terms in
\mbox{Eq.\ (\ref{eq:delvII})} and the relative velocities simply become
\begin{equation}
  \Delta V_{12}^2 = \Delta V_\mathrm{II}^2 = V_\mathrm{g}^2 \left( \frac{1}{1+\mathrm{St}_1} + \frac{1}{1+\mathrm{St}_2} \right).
  \label{eq:largelim}
\end{equation}
This result can, of course, directly be obtained from the $V_{\mathrm{p}i}$ terms (Eq.\
(\ref{eq:Vp3})) since the cross-term vanishes in this regime. For small
$\mathrm{St}_2$ relative velocities are still $\sim V_\mathrm{g}$; however, if both
Stokes numbers are large, the relative velocity decreases roughly with
the square root of the smallest particle stopping time. Note that the linear fit of
\citet{1984Icar...60..553W} in this regime (his Eq.\ (15)) is inappropriate (see, however, \citet{1980A&A....85..316V,1988mess.book..348W,1993prpl.conf.1031W,1993Icar..106..102C} in which a square-root fall off is advocated).
Since an explicit, closed-form solution to the \citet{1980A&A....85..316V} and MMV expressions for $\Delta V_{12}$ has not previously been
available, many dust coagulation models (e.g., \citet{2001ApJ...551..461S,2005A&A...434..971D,2007A&A...461..215O}) have relied on the
\citet{1984Icar...60..553W} fits to calculate relative velocities. 
Turbulent motions and relative velocities for particles in the $t_\mathrm{s} > t_\mathrm{L}$ regime have therefore been underestimated in these calculations. 
However, concerning these works, we also think no major conclusions have been affected, since the error is introduced only for large dust particles, that is, if the system is already well evolved.

\subsection{\label{sec:contour}Contour plots}
Following \citet{1980A&A....85..316V} and MMV we also
present our results as contour plots. Figure\ \ref{fig:contour}A shows, for comparison, the results of MMV, obtained by numerical evaluation of the integrals involved without an inner turbulent scale ($Re\rightarrow \infty$). The next three panels of Fig.\ \ref{fig:contour} show the result using our closed-form expressions derived from Eq.\ (\ref{eq:delv}). In panel B, the $y_a^*$ approximation has been used and, like Fig.\ 2 of MMV (panel A), the inner scale of the turbulence is extended to infinity so that Eqs.\ (\ref{eq:fullintermlim}, \ref{eq:largelim}) apply. 
Somewhat systematically higher values for $\Delta V_{12}$ when compared to MMV can be explained by the CH03 approximation for $V_\mathrm{p}$ (see Eq.\ (\ref{eq:Vp2})) but these discrepancies are less than $\sim$10\%. 
In panels C and D we show the contour plots corresponding to the other formulations for $k^*$ (see Fig.\ \ref{fig:tstarplot}), i.e., the exact solution for $y^*$ (panel C) and the CH03 empirical approximation (panel D).
The differences between these three methods for determining $k^*$ differ around the $\mathrm{St}=1$ point (see Fig.\ \ref{fig:tstarplot}) and are reflected in the contour plots. For $\mathrm{St} \approx 1$, panel C compares best to the numerical result of MMV, but no significant errors are made when using the $y_a^*$ approximation or the CH03 formula for $k^*$.  

In panels C and D of Fig.\ \ref{fig:contour}, a Reynolds number of $Re=10^8$ has been adopted. For $\mathrm{St} < 10^{-4}$, therefore, velocities are greatly suppressed since only class I eddies remain to generate relative velocities and relative velocities disappear completely for equal friction times. Also, the contours are much closer spaced since in this limit the velocity $\Delta V_{12}$ is proportional to $\mathrm{St}$ (see Eq.\ (\ref{eq:smalllim})).

\section{\label{sec:conclusions}Conclusions}
We have extended and, essentially, completed the work of \citet{2003Icar..164..127C}, 
who derived explicit, closed-form expressions for particle velocities in 
turbulence based on the physics originally developed by 
\citet{1980A&A....85..316V} and \citet{1991A&A...242..286M}. Within the
framework of this physics, the only approximations used here are in Eq.\ (\ref{eq:Vp2}) 
for the particle velocities (where {\it a posteriori} comparisons with exact numerical solutions indicate the approximation is well justified) and in Eqs.\ (\ref{eq:vreldef2}) {\it et seq} where the 
systematic velocity $V_\mathrm{o}$ is neglected to simplify calculating the boundary between eddy 
classes (generalizing this step should be straightforward, however). The full analytic
expression for $\Delta V_{12}$ is given by Eq.\ (\ref{eq:delv}) (or by the sum of
Eqs.\ (\ref{eq:delvI}, \ref{eq:delvII})), but more simple, explicit expressions apply in restricted 
regimes (provided $Re^{1/2} \gg 1$):
\begin{itemize}
  \item Equation\ (\ref{eq:verysmalllim}), in the very small particle limit
($t_1 \ll t_\eta$);
  \item Equation\ (\ref{eq:fullintermlim}), in the ``fully intermediate''
regime, i.e., for $t_\eta \ll t_1 \ll t_\mathrm{L}$;
  \item Equation\ (\ref{eq:largelim}), for $t_1 \ge t_\mathrm{L}$.
\end{itemize}
Near the $t_1 = t_\eta$ and $t_1 = t_\mathrm{L}$ turning points the behavior is more complex (see Fig.\ \ref{fig:delv}) and for accurate analytical approximations
one has to revert to the full expressions for $\Delta V$ given by \mbox{Eqs.\ (\ref{eq:delv}, \ref{eq:delvI}, \ref{eq:delvII})}.
\begin{acknowledgements}
  The authors thank Robert Hogan for computational assistance in creating Fig.\ \ref{fig:gh} and the referee, H. V\"olk, for positive feedback. CWO acknowledges support from the Netherlands Organisation for Scientific Research (NWO) that made this work possible. JNC's contributions were 
supported by a grant from NASA's Planetary Geology and Geophysics Program.
\end{acknowledgements}
\bibliographystyle{aa}
\bibliography{6899note}
\begin{appendix}
\section{\label{app:app1} A more accurate closed-form solution for $V_\textrm{p}$ and all related velocities, using power-law approximations to the functions $g$ and $h$}
\begin{figure}[t]
  \centering
  \includegraphics[width=88mm]{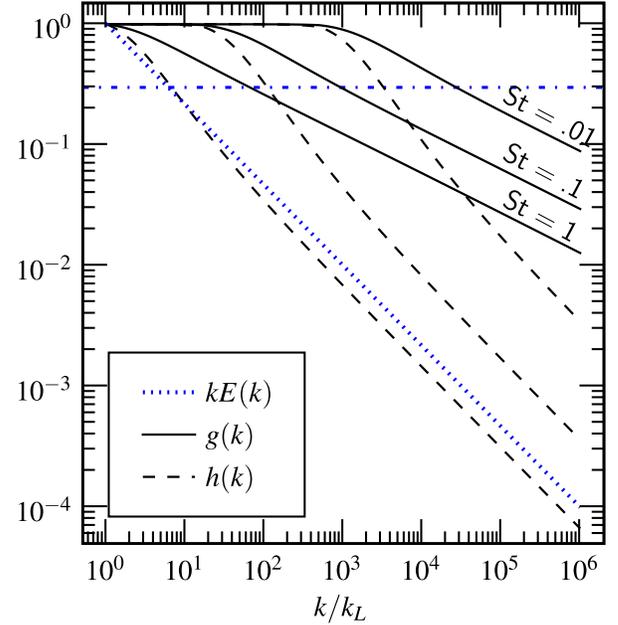}
  \caption{The functions $g$ (solid lines) and $h$ (dashed lines) plotted for three different Stokes numbers at a Reynolds number of $10^8$. After $k>k^*$ the functions show power-law behavior. The power spectrum (weighted by $k$) is plotted by the dotted line.}
  \label{fig:gh}
\end{figure}
In Sect.\ \ref{sec:definitions} the very simple approximation $g(\chi)=h(\chi)=1$ was introduced for all stokes numbers $\mathrm{St}$ and eddy scales $k$. It proves to be quite adequate for most purposes; however, as noted in Sect.\ \ref{sec:contour}, small inaccuracies remain at the 10\% level because the approximation overestimates the contributions of fast eddies to $V_\textrm{p}^2$ and other velocity components.  Figure\ \ref{fig:gh} shows the detailed behavior of the functions $g$ and $h$ for $Re=10^8$ for Stokes numbers of $\mathrm{St}$ = 0.01, 0.1, and 1.0. The inflection point for all three values of $\mathrm{St}$ is at $k=k^*$ (recall that $k^*/k_\mathrm{L} \approx 1 + \tfrac{1}{2}\mathrm{St}^{-3/2}$). For $k > k^*$, the functions are well approximated by power-laws of $-1/3$ and $-3/4$, respectively, i.e., $g(k)=(k/k^*)^{-1/3}$ and $h(k)=(k/k^*)^{-3/4}$. The success of the approximation of Sect.\ \ref{sec:definitions} is due to the fact that the power in the weighting function $k E(k)$ (dotted line; we multiply with $k$ since we compare logarithmically) decreases rapidly with increasing $k$; thus by the time the assumption $g(k)=h(k)=1$ becomes really bad, the relative contribution of successive terms has become small. For small $\mathrm{St}$, the weighted contribution of eddy power has already become very small even before $k \sim k^*$ (the logic of CH03). For $\mathrm{St} =1$ or larger, the weighting function has dropped by nearly an order of magnitude by the time $h(k)$ (the faster-decreasing function) has dropped to 0.3 (dashed-dotted line), and this seems to account for the success of our simple assumption. 

This behavior can be understood from the definition of $\chi = K t_k k V_\mathrm{rel}$. For $k \gg k^*$, $K \approx 1$ and $V_\mathrm{rel} \lesssim V_\mathrm{g}$ are both constant. Then, because $t_k \propto k^{-2/3}$, $\chi$ scales as $\chi \propto k^{1/3}$ and becomes large at large $k$. Since $g(\chi) = \arctan(\chi)/\chi \propto \chi^{-1}$ for large $\chi$, we get that $g(k) \propto k^{-1/3}$. Similarly, $h(\chi) \propto k^{-2/3}$, which is a bit shallower than the $-3/4$ exponent observed over most regions of interest (Fig\ \ref{fig:gh}). While the $-2/3$ exponent is reached at large $k$, the $-3/4$ exponent seems more appropriate at intermediate $k$. Yet, in our subsequent analysis, we will use the large-$k$ limit for this exponent ($-2/3$) because it simplifies the math. Thus, we approximate the $g$ and $h$ behavior as follows: unity for $k<k^*$, and power laws in $k/k^*$ with exponents of $-1/3$ and $-2/3$ for $k>k^*$. Then Eq.\ (\ref{eq:Vp0}) becomes 
\begin{multline}
  V_\mathrm{p}^2 = \int_{k_\mathrm{L}}^{k^*} dk\ 2E(k) \left( 1 - K^2 \right) \\
     + \int_{k^*}^{k_{\eta}} dk\ 2E(k) \left( 1-K \right)
\left[ \left(\frac{k}{k^*}\right)^{-1/3} + K  \left(\frac{k}{k^*}\right)^{-2/3} \right].
  \label{eq:app1}
\end{multline}
Where we have that $k_\eta \le k^* \le k_\mathrm{L}$ such that in the case of very small or very large particles one of the integrals vanishes (Sect.\ \ref{sec:tstar}). Since the approximation $g=h=1$ still holds for $k<k^*$ (or for $t>t^*$) the velocities resulting due to class 1 eddies (Eq.\ (\ref{eq:delvI})) are not affected; the new approximation only affects Eq.\ (\ref{eq:delvII}). By writing $K=\mathrm{St}/(\mathrm{St} + x^{-2/3})$, $E(k) = E_\mathrm{L} (k/k_\mathrm{L})^{-5/3} \propto x^{-5/3}$ with $x=k/k_\mathrm{L}$ the solution to Eq.\ (\ref{eq:app1}) involves integrals of the form
\begin{equation}
  \int dx\ x^{-5/3+p} \left( \frac{\mathrm{St}}{\mathrm{St} +x^{-2/3}} \right)^{n=[1,2]}
\end{equation}
with $n=2$ for the $K^2$ term and $p=-1/3$ or $-2/3$. These integrals can be solved analytically. Going to ``t-space'', however, gives somewhat cleaner solutions and we will from here on follow that approach and show how it affects relative velocities, i.e., $\Delta V_\mathrm{II}$. After the change of variables ($t_k/t^* = (k/k^*)^{-2/3}$) the second term of Eq.\ (\ref{eq:app1}) becomes
\begin{equation}
  \frac{V_g^2}{t_\mathrm{L}} \int_{t_\eta}^{t^*} dt_k\ \left( 1-K \right) \left( \frac{t_k}{t^*} \right)^{1/2} + (1-K)K \left( \frac{t_k}{t^*} \right).
  \label{eq:app3}
\end{equation}
We now introduce the dimensionless variable $y = t_k/t_\mathrm{s}$ (cf.\ Eq.\ (\ref{eq:vreldef2})). Then $t_k/t^* = y/y^*$ with $y^*=t^*/t_\mathrm{s}$. Also $K=1/(1+y)$ and $1-K=y/(1+y)$ and Eq.\ (\ref{eq:app3}) becomes
\begin{align}
  & \frac{V_g^2 t_\mathrm{s}}{t_\mathrm{L}} \int_{t_\eta/t_\mathrm{s}}^{t^*/t_\mathrm{s}} dy\ (y^*)^{-1/2} \frac{y^{3/2}}{1+y} + (y^*)^{-1} \frac{y^2}{(1+y)^2}  \\
  & = \frac{V_g^2 t_\mathrm{s}}{t_\mathrm{L}} \left\{ (y^*)^{-1/2} \Big[ I_h(y) \Big]^{t^*/t_\mathrm{s}}_{t_\eta/t_\mathrm{s}} + (y^*)^{-1}  \Big[ I_g(y) \Big]^{t^*/t_\mathrm{s}}_{t_\eta/t_\mathrm{s}} \right\}
\end{align}
in which the functions $I_h(y)$ and $I_g(y)$ are defined as
\begin{align}
  I_h(y) \equiv&\ \int_0^y dz\ \frac{z^{3/2}}{1+z} = \left(\frac{2}{3}y-2\right) \sqrt{y} + 2\arctan(\sqrt{y}) \\
  I_g(y) \equiv&\ \int_0^y dz\ \frac{z^2}{(1+z)^2} = \frac{(2+y)y}{1+y} - 2\log(1+y)
\end{align}
The expressions for $\Delta V_\mathrm{II}$ now consist of several contributions. First, $I_h(y)$ and $I_g(y)$ are evaluated at both the upper ($y^*$) and lower ($y_\eta$) limits. This must be done for both particles 1 and 2,
because the $\Delta V_\mathrm{II}$ term (Eq.\ (\ref{eq:delvII})) has separate contributions from each
particle. For the particle of highest friction time (say this is $t_1$) the power-law approximation for $g$ and $h$ holds over the range $\Delta V_\mathrm{II}$ is calculated, i.e., $t_s \le t_{1k} \le t_{12}^* = t_1^*$. However, for the second particle the power-law approximation only holds for $t_{2k} \le t_2^*$, while for the remaining range over which the integral in $\Delta V_\mathrm{II}$ is evaluated, i.e., $t^*_2 \le t_{2k} \le t_1^*$, the $g=h=1$ approximation applies.

This gives us several terms that contribute to $\Delta V_\mathrm{II}$. Collecting these terms, the new expression for $\Delta V_\mathrm{II}$ becomes
\begin{align}
  \nonumber
  \Delta V_\mathrm{II} =\ \frac{V_g^2}{t_\mathrm{L}}& \left\{t_1 \left( \frac{t_1^*}{t_1} \right)^{-1/2} \bigg[ I_h(y) \bigg]_{t_\eta/t_1}^{t_1^*/t_1} + (1\leftrightarrow2) \right. \\
  \nonumber
                                           &\; + t_1 \left( \frac{t_1^*}{t_1} \right)^{-1}   \bigg[ I_g(y) \bigg]_{t_\eta/t_1}^{t_1^*/t_1} + (1\leftrightarrow2) \\
 &\; + \left.\left[ t_k + \frac{t_2^2}{t_2 + t_k} \right]^{t_1^*}_{t_2^*} \right\},\qquad (t_1 \ge t_2).
 \label{eq:delvIIimproved}
\end{align}
Although still fully analytical, this more accurate expression for $\Delta V_\mathrm{II}$ is also more complicated and we did not present it in the main body of the paper. Equation\ (\ref{eq:delvIIimproved}) is useful, however, for readers whose applications demand this higher level of accuracy. 

\section{\label{ap:app2}Derivation of Eq.\ (\ref{eq:fullintermlim})}
We consider the limiting case of $t_\eta \ll t_1 \ll t_\mathrm{L}$. The $y_a^*$ approximation for $\mathrm{St}^*_{12}$ then holds, i.e., $\mathrm{St}_{21}^* \approx y_a^* \mathrm{St}_1$ with $y_a = 1.6$. We will now argue that we can neglect the $Re^{-1/2}$ terms in Eq.\ (\ref{eq:delvII}). For particle 1 this is obvious since $\mathrm{St}_1 \gg Re^{-1/2}$. The last term (where $Re^{-1/2}$ is in the denominator) then becomes simply $-\mathrm{St}_1$. However, for the interchange term a similar approximation
\begin{equation}
   \frac{\mathrm{St}_2^2}{\mathrm{St}_2 + Re^{-1/2}} \approx \mathrm{St}_2, \qquad 
   \label{eq:interterm}
\end{equation}
is not that obvious since we have not put a constraint on $\mathrm{St}_2$. For example, if $\mathrm{St}_2 \ll Re^{-1/2}$ the $Re^{-1/2}$ term dominates the denominator. However, in that case this term \textit{and} its approximation are small anyway compared to $-\mathrm{St}_1$, such that by making the approximation in Eq.\ (\ref{eq:interterm}) our final result is not affected. Similarly, if $\mathrm{St}_2 \sim Re^{-1/2}$, Eq.\ (\ref{eq:interterm}) (which goes to $\sim \tfrac{1}{2}Re^{-1/2}$) or its approximation ($\sim Re^{-1/2}$) are insignificant since $\mathrm{St}_1 \gg Re^{-1/2}$. Only if $\epsilon \sim 1$, i.e., $\mathrm{St}_2 \gg Re^{-1/2}$, does the $\mathrm{St}_2$ term matter, but then the approximation in  Eq.\ (\ref{eq:interterm}) is well justified. All terms in Eq.\ (\ref{eq:delvII}) are then linear in Stokes and we can reduce it to
\begin{equation}
  \frac{\Delta V_\mathrm{II}^2}{V_\mathrm{g}^2} = \left( 2y_\mathrm{a}^* - (1-\epsilon) + \frac{1}{1+y_\mathrm{a}^*} + \frac{\epsilon^2}{\epsilon + y_\mathrm{a}^*} \right) \mathrm{St}_1,
  \label{eq:delvIIred}
\end{equation}
with $\epsilon = \mathrm{St}_2/\mathrm{St}_1 \le 1$. Similarly, Eq.\ (\ref{eq:delvI}) becomes
\begin{equation}
  \frac{\Delta V_\mathrm{I}^2}{V_\mathrm{g}^2} = \frac{1-\epsilon}{1+\epsilon} \left(  \frac{1}{y_a^* + 1} - \frac{\epsilon^2}{y_a^* + \epsilon} \right) \mathrm{St}_1.
  \label{eq:delvIred}
\end{equation}
Combining these expressions and collecting the $1/(1+y_a^*)$ and $\epsilon^2/(y_a^*+\epsilon)$ terms then gives Eq.\ (\ref{eq:fullintermlim}).
\end{appendix}
\end{document}